\documentclass{article}

\newcounter{lista}
\title{Painlev\'e Analysis in Superspace}
\author{Alin A. Constandache\thanks{e-mail: alexc@pas.rochester.edu}\\Department of Physics and Astronomy\\University of Rochester}

\begin{document}
\bibliographystyle{h-physrev}

\maketitle
\begin{abstract}
A method for carrying out the Painlev\'e test in superspace is proposed. The method is then applied to the one-parameter $N=1$ supersymmetric extensions of the KdV equation.
\end{abstract}
\vfill\pagebreak

\section{Introduction}
The Painlev\'e analysis is a simple and useful tool for testing the
integrability of ordinary differential equations (ODEs) and partial
differential equations (PDEs) by analyzing the singularities of the
solutions. It has its origins in the works of S. Kowalewski,
P. Painlev\'e and B. Gambier \cite{Kovalevski:Painleve,Painleve:Painleve,Gambier:Painleve}, who introduced it in the study of ODEs. 
The test has since been extended to PDEs by Ablowitz, Ramani and Segur
\cite{Abloetal:Painleve1,Abloetal:Painleve2} who conjectured that any similarity
reduction to an ODE, of a given PDE that is integrable by the
method of inverse scattering, must have the Painlev\'e property. 
Weiss, Tabor and Carnevale \cite{Weissetal:Painleve} have proposed a
version of the test that can be applied directly to PDEs, without any
reduction to ODEs. This has been explicitly verified in the case of a
number of integrable systems such as the Burgers' equation, the KdV
equation, the Boussinesq equation, the KP equation and so on.

More recently, there has been interest in the study of supersymmetric
integrable systems for a variety of reasons. In addition to the usual
bosonic dynamical variables, such models also contain anti-commuting
Grassmann variables. The Painlev\'e property of such systems have been
studied only in a handful of cases and there in component formalism. The
presence of fermions, in such systems,  gives rise to some new
properties, mainly in the recursion of the coefficient functions. The
natural manifold on which a supersymmetric system is defined is a
superspace, which contains Grassmann coordinates in addition to the
standard  bosonic coordinates. The study of the Painlev\'e property of
a supersymmetric integrable system should naturally be carried out on
this supermanifold. So far, however, a description of the Painlev\'e
analysis in superspace is lacking \cite{Mathieu:Open}. In this paper, we make an attempt
at generalizing the Painlev\'e analysis to superspace. In section {\bf
\ref{secSusyP}}, we describe the general formalism of how the Painlev\'e analysis can
be carried out in superspace. In section {\bf \ref{secSusyKdV}}, we apply our method to study the integrability of the $N=1$ supersymmetric KdV system and we 
present a brief conclusion in section {\bf \ref{secConcl}}.

\section{Painlev\'e Analysis in Superspace}\label{secSusyP} 
The Painlev\'e analysis, for standard bosonic systems, consists of
expanding the solution as a power series
$$
u = \sum_{k=0}^{\infty} u_{k}\rho^{k-\beta}
$$
where
$$
\rho = 0
$$
defines the singularity manifold of the system of solutions. The PDE
is said to have the Painlev\'e property if the solution is single
valued about the movable singularity manifolds. This can happen only if
the exponent $\beta$ is an integer and that the coefficients
$u_{k}$'s are related by some recursion relation (to be determined
from the dynamical equation) such that the solution is analytic near the
singularity manifold. A naive generalization of this method to
superspace would consist of expanding the solution (which would be a
superfield consisting of both bosonic and fermionic variables) of the
dynamical equation in superspace, in the form
\begin{equation}\label{eqnpainleveexp}
\Phi(t,x,\theta) = \sum_{k=0}^{\infty} \Phi_{k}\rho^{k-\beta}\label{1}
\end{equation}
where, for simplicity, we are considering a superspace with a single
Grassmann variable $\theta$. The superfield $\Phi$, if it is
fermionic, will have an expansion of the form
\begin{equation}\label{eqnsuperfield}
\Phi(t,x,\theta) = \psi(t,x) + \theta u(t,x)\label{2}
\end{equation}
where $\psi$ and $u$ are respectively the fermionic and the bosonic
dynamical variables in the component form. The singularity manifold in
(\ref{eqnpainleveexp}) would now be a supermanifold and the coefficient
functions in (\ref{eqnsuperfield}) would, in general, depend on the Grassmann
coordinates of the manifold. A simple extrapolation of the results for
bosonic systems would then say that the superspace equation would have
the Painlev\'e property provided $\beta$ is an integer and that we can
find recursion relations between the coefficient functions such that
the solution is analytic near the singularity supermanifold. A little
thought, however, easily convinces one that such a generalization is
doomed to fail for an obvious reason. On this superspace,
there is a covariant derivative
\begin{equation}
D = {\partial\over \partial\theta} + \theta {\partial\over \partial
x}\label{3}
\end{equation}
which is fermionic and is invariant under a supersymmetry
transformation (translations of the superspace) satisfying
\begin{equation}
D^{2} = {\partial\over \partial x}\label{4}
\end{equation}
The dynamical equations in superspace, therefore, in general, admit
covariant derivatives acting on the superfield variables, in addition
to the usual bosonic derivative with respect to $x$. As a result, it
becomes impossible to obtain a recursion directly between the
coefficients $\Phi_{k}$ in (\ref{eqnpainleveexp}). We want to emphasize here
that, for systems which do not explicitly involve covariant
derivatives, the Painlev\'e analysis will go through with the expansion of the form (\ref{eqnpainleveexp}). However, it would be more useful to have a general description which works in all cases.

The solution to this problem is, in fact, quite simple. Let us, for
simplicity, consider a fermionic superfield $\Phi(x,\theta)=\psi(x)+\theta\,u(x)$ and an equation of motion which has the form
\begin{equation}\label{eqnsusygen}
\Phi_{t}=P(\Phi_{(k)},\;(D\,\Phi)_{(k)})
\end{equation}
where $P(\Phi_{(k)},\;(D\,\Phi)_{(k)})$ is a polynomial in $\Phi$, $(D\,\Phi)$ and their $x$-derivatives up to some order (we use the condensed notation $A_{(k)}\equiv\partial^{k}A$). Taking the covariant derivative of (\ref{eqnsusygen}) will give us an additional equation of the form
\begin{equation}\label{eqnsusygenderiv}
(D\,\Phi)_{t}=Q(\Phi_{(k)},\;(D\,\Phi)_{(k)})
\end{equation}
where $Q$ is another polynomial in $\Phi$, $(D\,\Phi)$ and their $x$-derivatives up to some order. Let $U$ now be a bosonic field independent of $\Phi$ and consider the system of coupled PDEs:
\begin{eqnarray}\label{eqnsusygenassoc}
\Phi_{t}&=&P(\Phi_{(k)},\;U_{(k)})\nonumber\\
&&\\
U_{t}&=&Q(\Phi_{(k)},\;U_{(k)})\nonumber
\end{eqnarray}
where $P$ and $Q$ are the same polynomials that appear on the right hand side  in (\ref{eqnsusygen}) and (\ref{eqnsusygenderiv}) only this time in variables $\Phi$ and $U$ (instead of $\Phi$ and $(D\,\Phi)$). This system has a few nice properties which make it interesting for our purpose:
\begin{list}{(\roman{lista})}{\usecounter{lista}}
\item it does not explicitly contain the covariant derivative, hence we can perform the Painlev\'e test on it in the usual way;
\item if $\Phi$ is a solution of (\ref{eqnsusygen}) then $(\Phi,\;(D\,\Phi))$ is a solution of (\ref{eqnsusygenassoc});
\item if every solution of (\ref{eqnsusygenassoc}) is analytic in a neighborhood of the singularity manifold $\rho$ and admits a Painlev\'e expansion in this neighborhood, then the same holds for every solution of (\ref{eqnsusygen});
\end{list}
Based on these observations we propose the following prescription for performing the Painlev\'e test in superspace:
\begin{list}{{\bf Step\ \arabic{lista}.}}{\usecounter{lista}}
\item Obtain the coupled system (\ref{eqnsusygenassoc}) associated with the
given PDE. 
\item Treating $U$ and $\Phi$ as independent superfields, look for
solutions of the form 
$$
\begin{array}{ll}
U=\sum_{k=0}^{\infty}U_{k}\rho^{k-\alpha},&\Phi=\sum_{k=0}^{\infty}\Phi_{k}\rho^{k-\beta}
\end{array}
$$
where $U_{k}$ are bosonic superfields, $\Phi_{k}$ are fermionic
superfields and  $\rho$ is another bosonic superfield representing the
singularity manifold. 
\item Find the integer values of $\alpha$ and $\beta$ for which the
system has enough resonances. 
\item Out of the cases found in the previous step, select the ones for
which all the compatibility conditions are satisfied. These are the
cases that will pass the test. 
\end{list}

The resulting calculations can be very tedious, therefore one might want to use a simplification known in the literature as the {\em Kruskal ansatz}. For a bosonic system, the Kruskal ansatz consists of looking for solutions of the form
$$
u(x,t)=\sum_{k=0}^{\infty}u_{k}(t)\rho^{k-\beta}(x,t)
$$
where $\rho(x,t)=x-\phi(t)$ with $\rho(x,t)=0$ representing the equation of the singularity manifold. The fact that for some $\phi$ the singularity manifold has this special form is guaranteed by the implicit function theorem provided that $\rho_{x}\not=0$. With this in mind it is easy to see that the Kruskal ansatz generalizes to superspace as follows: for a system of the form (\ref{eqnsusygenassoc}) look for solutions $\Phi=\sum_{k=0}^{\infty}\Phi_{k}\rho^{k-\beta}$, $U=\sum_{k=0}^{\infty}U_{k}\rho^{k-\alpha}$ with restrictions $\rho_{x}=1$, $\Phi_{k\,x}=0$, $U_{k\,x}=0$.

We turn our attention now to the connection that exists between the covariant version of the Painlev\'e analysis, as described in the previous paragraphs, and the traditional way of carrying it out in components. Let us consider the most general monomial in $\Phi, (D\Phi)$ and their $x$-derivatives:
\begin{equation}\label{eqngenmonom}
\prod_{k=0}^{n}[\Phi_{(k)}^{p_{k}}\,(D\,\Phi_{(k)})^{q_{k}}]
\end{equation}
where $A_{(k)}\equiv\partial^{k}A$, $q_{k}$ are positive integers and $p_{k}\in\{0, 1\}$. Writing this in components we get:
\begin{eqnarray}
&&\prod_{k=0}^{n}[\Phi_{(k)}^{p_{k}}\,(D\,\Phi_{(k)})^{q_{k}}]=\nonumber\\
&=&\prod_{k=0}^{n}[(\psi_{(k)}^{p_{k}}+\theta\,p_{k}\psi_{(k)}^{p_{k}-1}u_{(k)})(u_{(k)}^{q_{k}}+\theta\,q_{k}u_{(k)}^{q_{k}-1}\psi_{(k+1)})]=\nonumber\\
&=&\prod_{k=0}^{n}[\psi_{(k)}^{p_{k}}u_{(k)}^{q_{k}}]+\theta\sum_{k=0}^{n}(-1)^{\sigma_{k}}(\psi_{(0)}^{p_{0}}u_{(0)}^{q_{0}})\dots(\psi_{(k)}^{p_{k}}u_{(k)}^{q_{k}})^{\clubsuit}\dots(\psi_{(n)}^{p_{n}}u_{(n)}^{q_{n}})\nonumber
\end{eqnarray}
where $\sigma_{k}=\sum_{j=0}^{k-1}p_{j}$ and \mbox{$(\psi_{(k)}^{p_{k}}u_{(k)}^{q_{k}})^{\clubsuit}=p_{k}u_{(k)}^{q_{k}+1}\psi_{(k)}^{p_{k}-1}+(-1)^{p_{k}}q_{k}u_{(k)}^{q_{k}-1}\psi_{(k)}^{p_{k}}\psi_{(k+1)}$}. Hence the components of (\ref{eqngenmonom}) are
\begin{equation}
\left\{\begin{array}{ll}
\prod_{k=0}^{n}[\psi_{(k)}^{p_{k}}u_{(k)}^{q_{k}}]\\
\sum_{k=0}^{n}(-1)^{\sigma_{k}}(\psi_{(0)}^{p_{0}}u_{(0)}^{q_{0}})\dots(\psi_{(k)}^{p_{k}}u_{(k)}^{q_{k}})^{\clubsuit}\dots(\psi_{(n)}^{p_{n}}u_{(n)}^{q_{n}})
\end{array}\right.
\end{equation}

If we compute the covariant derivative of (\ref{eqngenmonom}) we obtain:
\begin{eqnarray}\label{eqnpartner1}
&&D\,\left(\prod_{k=0}^{n}(\Phi_{(k)}^{p_{k}}\,(D\,\Phi_{(k)})^{q_{k}}\right)=\nonumber\\
&=&\sum_{k=0}^{n}(-1)^{\sigma_{k}}(\Phi_{(0)}^{p_{0}}(D\,\Phi_{(0)})^{q_{0}})\dots D\,(\Phi_{(k)}^{p_{k}}(D\,\Phi_{(k)})^{q_{k}})\dots(\Phi_{(n)}^{p_{n}}(D\,\Phi_{(n)})^{q_{n}})=\nonumber\\
&=&\sum_{k=0}^{n}(-1)^{\sigma_{k}}(\Phi_{(0)}^{p_{0}}(D\,\Phi_{(0)})^{q_{0}})\dots (\Phi_{(k)}^{p_{k}}(D\,\Phi_{(k)})^{q_{k}})^{\clubsuit}\dots(\Phi_{(n)}^{p_{n}}(D\,\Phi_{(n)})^{q_{n}})
\end{eqnarray}
Denoting $(D\,\Phi)$ by $U$ in (\ref{eqngenmonom}) and (\ref{eqnpartner1}) we obtain two polynomials which have the same structure as the components of (\ref{eqngenmonom}). Due to the linearity of $D$, this property extends to arbitrary polynomials, which leads to the following result: given a supersymmetric PDE of the form (\ref{eqnsusygen}) one can associate with it a system of coupled PDEs (\ref{eqnsusygenassoc}) as previously describred. If the form of (\ref{eqnsusygen}) in components is
\begin{eqnarray}\label{eqncompsys}
\psi_{t}&=&p(\psi_{(k)}, u_{(k)})\nonumber\\
&&\\
u_{t}&=&q(\psi_{(k)}, u_{(k)})\nonumber
\end{eqnarray}
then the polynomials $P$ and $p$ have identical structure and so do $Q$ and $q$. This means that the initial PDE passes the covariant version of the Painlev\'e test if and only if it passes the traditional componentwise version of the test.

Finally, we note that a similar argument can be made if the superfield
we start with is a bososnic superfield.

\section{Application to $N=1$ susy-KdV}\label{secSusyKdV}
In this section we apply the superspace Painlev\'e analysis, as defined in section \ref{secSusyP}, to the $N=1$ supersymmetric
extensions of the KdV equation with one arbitrary parameter  and we
regain the known result that for only two values of the parameter
($c=0$ and $c=3$) the system is integrable. 

The one-parameter family of supersymmetric extensions has the form:\nocite{Mathieu:susyKdV}
\begin{equation}
\Phi_{t}=\Phi_{xxx}+(6-c)\,(D\Phi)\,\Phi_{x}+c\,\Phi\,(D\Phi)_{x}
\end{equation}
therefore the coupled system we will analyze is
\begin{eqnarray}\label{eqnsusyKdV}
\Phi_{t}&=&\Phi_{xxx}+(6-c)\,U\,\Phi_{x}+c\,\Phi\,U_{x}\nonumber\\
&&\\
U_{t}&=&U_{xxx}+6\,U\,U_{x}-c\,\Phi\,\Phi_{xx}\nonumber
\end{eqnarray}
Note that this no longer involves the covariant derivative explicitly. For simplicity we will use the Kruskal ansatz, namely we will look for
expansions of the form 
\begin{equation}
\begin{array}{ll}U=\sum_{k=0}^{\infty}U_{k}\rho^{k},&\Phi=\sum_{k=0}^{\infty}\Phi_{k}\rho^{k}\end{array}
\end{equation}
with the restrictions $U_{kx}=0$, $\Phi_{kx}=0$,
$\rho_{x}=1$. Plugging this back into (\ref{eqnsusyKdV}) we get: 
\begin{eqnarray}\label{eqnseries}
\sum_{k=0}^{\infty}\Phi_{kt}\,\rho^{k}&+&\sum_{k=0}^{\infty}(k-\beta)\,\Phi_{k}\,\rho_{t}\,\rho^{k-1}=\nonumber\\
&=&\sum_{k=0}^{\infty}(k-\beta)(k-\beta-1)(k-\beta-2)\,\Phi_{k}\,\rho^{k-3}+\nonumber\\
&+&(6-c)\,\rho^{-\alpha}\,\sum_{k=0}^{\infty}\,\sum_{l=0}^{k}\,(l-\beta)\,U_{k-l}\,\Phi_{l}\,\rho^{k-1}+\nonumber\\
&+&c\,\rho^{-\alpha}\,\sum_{k=0}^{\infty}\,\sum_{l=0}^{k}\,(l-\alpha)\,\Phi_{k-l}\,U_{l}\,\rho^{k-1}\nonumber\\
&&\\
\sum_{k=0}^{\infty}\,U_{kt}\,\rho^{k}&+&\sum_{k=0}^{\infty}\,(k-\alpha)\,U_{k}\,\rho_{t}\,\rho^{k-1}=\nonumber\\
&=&\sum_{k=0}^{\infty}\,(k-\alpha)(k-\alpha-1)(k-\alpha-2)\,U_{k}\,\rho^{k-3}+\nonumber\\
&+&6\,\rho^{-\alpha}\,\sum_{k=0}^{\infty}\,\sum_{l=0}^{k}\,(l-\alpha)\,U_{k-l}\,U_{l}\,\rho^{k-1}-\nonumber\\
&-&c\,\rho^{-(2\,\beta-\alpha)}\,\sum_{k=0}^{\infty}\,\sum_{l=0}^{k}\,(l-\beta)(l-\beta-1)\,\Phi_{k-l}\Phi_{l}\,\rho^{k-2}\nonumber
\end{eqnarray}

In order for the system to pass the test, it must have six resonances, hence we must have a polynomial of order six in $k$ in the determinant of the recursion matrix. This is possible only if $\alpha=2$ and $\beta=2$. 

Equating coefficients in (\ref{eqnseries}) we get
\begin{eqnarray}\label{eqncoef}
\Phi_{k-3\,t}+(k-4)\,\Phi_{k-2}\,\rho_{t}&=&(k-2)(k-3)(k-4)\,\Phi_{k}+\nonumber\\
&+&(6-c)\,\sum_{l=0}^{k}\,(l-2)\,U_{k-l}\,\Phi_{l}+\nonumber\\
&+&c\,\sum_{l=0}^{k}\,(l-2)\,\Phi_{k-l}\,U_{l}\nonumber\\
&&\\
U_{k-3\,t}+(k-4)\,U_{k-2}\,\rho_{t}&=&(k-2)(k-3)(k-4)\,U_{k}+\nonumber\\
&+&3\,(k-4)\,\sum_{l=0}^{k}U_{k-l}\,U_{l}-\nonumber\\
&-&c\,(k-4)\,\sum_{l=0}^{k+1}\,l\Phi_{k+1-l}\,\Phi_{l}\nonumber
\end{eqnarray}
In particular, for $k=0$, this becomes:
\begin{eqnarray}
\Phi_{0}\,(U_{0}+2)&=&0\nonumber\\
&&\\
-3\,U_{0}\,(2+U_{0})+c\,\Phi_{0}\,\Phi_{1}&=&0\nonumber
\end{eqnarray}
There are only two ways to satisfy these two equations simultaneously:
\begin{list}{(\roman{lista})\ }{\usecounter{lista}}
\item $U_{0}=-2$ and $\Phi_{0}\,\Phi_{1}=0$
\item $U_{0}=-2$ and $c=0$
\end{list}
The last of the two cases corresponds to the susy-KdVB equation and it also arises as a special subcase of the first, as we will see later. Therefore, without any loss of generality we can restrict our
analysis to the first case. The recurrence then takes the form: 
\begin{equation}
\left(\begin{array}{cc}
k\,(k^{2}-9\,k+14+2\,c)&(c\,k-12)\,\Phi_{0}\\
-c(k-4)(k-1)\,\Phi_{1}&(k-4)(k+1)(k-6)
\end{array}\right)
\left(\begin{array}{c}\Phi_{k}\\U_{k}\end{array}\right)=
\left(\begin{array}{c}F_{k}\\G_{k}\end{array}\right)
\end{equation} 
where $F_{k}$ depends on
$U_{0},\dots,U_{k-1},\Phi_{0},\dots,\Phi_{k-1}$ and their derivatives,
while $G_{k}$ depends on
$U_{0},\dots,U_{k-1},\Phi_{0},\dots,\Phi_{k-1},\Phi_{k+1}$ and their
derivatives. 

The resonances are therefore given by the roots of the polynomial
$$
k\,(k-4)(k+1)(k-6)(k^{2}-9\,k+14+2\,c)
$$
and in order to have six of them, the quadratic factor must have two
integer roots, both greater or equal to $-1$. This restriction leaves
us with only five possible values for $c$: 
\begin{list}{(\roman{lista})}{\usecounter{lista}}
\item $c=3$.

The system has resonances at $-1, 0, 4, 5, 6$ and the resonance at
$k=4$ has multiplicity 2. We checked that all the compatibility
conditions for this case are satisfied and therefore it passes the
test, as expected. The arbitrary functions are $\rho,
\Phi_{0},\Phi_{4},U_{4},\Phi_{5},U_{6}$. 
\item $c=2$

The resonances occur at $-1, 0, 3, 4, 6$ and the resonance $k=6$ has
multiplicity 2. However, the compatibility condition for $k=3$ is not
satisfied. 
\item $c=0$.

This is the susy-KdVB case that we have mentioned before. The
resonances are $-1, 0, 2, 4, 6, 7$. All the compatibility conditions
are satisfied, thus this case also passes the test. The arbitrary
functions are $\rho, \Phi_{0}, \Phi_{2}, U_{4}, U_{6}, \Phi_{7}$. 
\item $c=-3$

The resonances are $-1, 0, 1, 4, 6, 8$. The compatibility condition at
$k=6$ does not hold. 
\item $c=-7$

The resonances are $-1, 0, 4, 6, 9$, with the one at $k=0$ having
multiplicity 2. However, the compatibility condition for $k=0$ is not
satisfied. 
\end{list}
Therefore, out of the five possible cases, only the two that are known
to be integrable have the Painlev\'e property.

\section{Conclusion}\label{secConcl}
We have generalised the Painlev\'e analysis to superspace by introducing an additional superfield such that all explicit occurences of the covariant derivative in the resulting equations of motion are eliminated. The Painlev\'e analysis can then be carried out for the resulting system. We have applied the method to the $N=1$ extensions of the KdV equation and regained the known result that only two of these extensions have the Painlev\'e property. However, unlike earlier analysis \cite{Mathieu:nonsusyPainleve}, here we work manifestly in superspace. The application of our method to $N=2$ supersymmetric systems is under investigation and will be reported in the future.

\section*{Acknowledgements}
This work has been supported in part by the U.S. Deptartment of Energy Grant \mbox{DE-FG 02-91ER40685}.  

\bibliography{main}

\end{document}